\documentstyle[aps,twocolumn,epsfig,floats]{revtex}

\begin{document}  
\draft
\wideabs{  
\author{Christof Gattringer}
\address{\medskip Institut f\"ur Theoretische Physik, Universit\"at
Regensburg, D-93040 Regensburg, Germany} 
\date{October 1, 2002}
\title{Calorons, instantons and constituent monopoles in SU(3) lattice 
gauge theory}
\maketitle
   
\begin{abstract}  
We analyze the zero-modes of the Dirac operator in quenched SU(3) 
gauge configurations at non-zero temperature and compare 
periodic and anti-periodic temporal boundary 
conditions for the fermions. It is demonstrated that for the different 
boundary conditions often the modes are localized at different 
space-time points and have different sizes. Our observations are 
consistent with patterns expected for Kraan - van Baal solutions of the 
classical Yang-Mills equations.
These solutions consist of constituent monopoles and the zero-modes 
are localized on different constituents for different boundary conditions. 
Our findings indicate that the excitations of the QCD vacuum
are more structured than simple instanton-like lumps.
\end{abstract}
 }

Understanding the relevant excitations of the QCD vacuum is a major 
scientific goal and lattice calculations can contribute valuable 
non-perturbative information. A prominent candidate for such relevant 
excitations are instantons which are particularly successful in 
providing a mechanism for chiral symmetry breaking \cite{diakonov}. 
Lattice calculations have helped to corroborate these ideas. Recently
several studies of the zero-modes and near zero-modes 
of the Dirac operator \cite{Dvecs,horvath1,selfdual} 
which provide a powerful and natural filter retaining the infrared 
fluctuations of the underlying gauge field were presented. 
It was established that 
the vacuum has a lumpy structure and the lumps are locally chiral 
\cite{horvath1} and (anti) self-dual \cite{selfdual}. 

However, it is generally believed, that instantons are not responsible 
for confinement. On the other hand all lattice calculations show that 
the deconfinement transition and the restoration of chiral symmetry 
coincide at a single critical temperature which suggests that there is 
an underlying mechanism linking the two phenomena. 
In recent years Kraan and van Baal \cite{kvb1}
have constructed new solutions for the classical SU(N) Yang Mills equations 
with compactified time, i.e.~at finite temperature (KvB solutions). 
The KvB solutions generalize the caloron solution \cite{HaSh} and
a lump of topological charge $1$ is built 
of N constituent monopoles, a feature
which might provide a link to the confinement 
problem. Note that for KvB solutions the monopole lumps appear in 
multiples of N. 
A single lump is not an independent object and only N lumps together
have topological charge $\pm 1$.
Strong evidence for SU(2) KvB solutions in cooled lattice 
gauge configurations were given 
for twisted \cite{kvblat1} and periodic boundary conditions 
\cite{kvblat2}. 
The zero-mode for the KvB solution \cite{kvb2} shows characteristic features 
which should allow to use the eigenvectors of the Dirac operator 
as a filter retaining 
infrared fluctuations and search for traces of KvB solutions
also in thermalized configurations. 

Before presenting our lattice calculations we briefly 
discuss a few properties of SU(3) KvB solutions and their zero-mode. 
KvB solutions \cite{kvb1} 
are characterized by the Polyakov loop at spatial infinity
which in a suitable gauge can be expressed as
$P_\infty = \exp (i 2\pi \mbox{diag} (\mu_1,\mu_2,\mu_3))$
with $\mu_1 + \mu_2 + \mu_3 = 0$ and 
$\mu_1 \le \mu_2 \le \mu_3 \le \mu_4 \equiv 1 + \mu_1$. 
The action density can be written down in a relatively simple form \cite{kvb1}
and depends in addition to the phases $\mu_1,\mu_2,\mu_3$ also on three
3-vectors $\vec{y}_1, \vec{y}_2, \vec{y}_3$. As remarked 
the KvB solutions can be seen to be superpositions of constituent 
monopoles and the 3 corresponding lumps are in space located 
at $\vec{y}_1, \vec{y}_2, \vec{y}_3$.
The $m$-th constituent monopole can be assigned a mass 
$8\pi^2(\mu_{m+1} - \mu_{m})$ (both the masses and the monopole 
positions $\vec{y}_i$ are in units of the inverse temperature).
A general choice for the parameters of
the KvB solution will give rise to three lumps at different positions with
different width. 

The zero-mode $\psi$ in the background of a KvB solution 
has been computed \cite{kvb2}
for arbitrary temporal boundary conditions $\psi(t + 1/T, \vec{x}) =
\exp(i 2\pi z) \psi(t, \vec{x})$ with $z = 1/2$ corresponding to anti-periodic
boundary conditions (b.c.) and $z = 0$ giving the periodic case. The remarkable
feature of the zero-mode is that it traces only one of the 
monopoles. In particular it is located on the $m$-th monopole
when $z \in \{\mu_m,\mu_{m+1}\}$. At least for well separated monopoles 
the width of the lump seen by the zero-mode 
depends on the size of the underlying gluonic lump and also the quantity
$\sigma = \min[z-\mu_m,\mu_{m+1}-z]$. The lump is narrow for large $\sigma$
and broad for small $\sigma$. This is an important feature in the deconfined
phase where due to the $Z_3$ symmetry of the action the Polyakov loop
can have three different phases $0, \pm 2\pi/3$. At sufficiently high 
temperatures configurations with real 
Polyakov loop have $\mu_i \sim 0, i = 1,2,3$. The anti-periodic zero-mode has
$z = 1/2$, is located on monopole 3 and has 
$\sigma \sim 1/2$ and thus is quite localized. The periodic zero-mode 
with $z = 0$ can be located on any of the monopoles. It has $\sigma \sim 0$ 
and is very broad. For configurations with phase $2\pi/3$ of the Polyakov loop
(the case with phase $-2\pi/3$ is similar)
one has $(\mu_1,\mu_2,\mu_3) \sim (-2/3,1/3,1/3)$. The anti-periodic
mode has $\sigma \sim 1/6$ while the periodic mode has $\sigma \sim 1/3$
and both of them are localized on monopole 1. This
implies that for complex Polyakov loop the roles are reversed and here the 
periodic mode is more localized. However, the difference in localization
is not as pronounced since the values $\sigma \sim 1/6$ and 1/3
differ less than $\sigma \sim 1/2$ and 0 obtained for real Polyakov loop. 

To summarize, if KvB solutions are present
in thermalized SU(3) gauge configurations then one should find instances
where the zero-modes with anti-periodic b.c.~and periodic b.c.~are located 
at different positions and the lumps they see can differ in size. In the 
deconfined phase the differences in size should show a particular pattern  
when comparing configurations with real and complex Polyakov loop. These are 
simple stable signatures which distinguish KvB solutions from instantons 
or calorons. 

In our numerical analysis 
we use quenched SU(3) gauge configurations generated with the 
L\"uscher-Weisz action \cite{LuWeact} on $6\times 20^3$ lattices. 
We adjust the inverse gauge coupling $\beta$ such that one ensemble is below
the critical temperature and the other one in the deconfined high 
temperature phase.
The lattice spacing $a$ and the values for the
temperature $T$ and $T_c$ for the L\"uscher-Weisz 
action listed in Table \ref{rundat} were determined in \cite{scales}.
\begin{table}[t]
\vspace{-5mm}
\begin{center}
\begin{tabular}{cccccc}
size & $\beta$ & statistics & $a$ [fm] & $T$ [MeV] & $T/T_c$ \\
$6\times20^3$ & 8.20 & 70 & 0.115 & 287 & 0.96 \\
$6\times20^3$ & 8.45 & 64 + 25 & 0.094 & 350 & 1.17 
\end{tabular}
\end{center}
\caption{Parameters of our calculations.
\label{rundat}}
\end{table}

We calculate the eigenvalues and eigenvectors of 
the chirally improved Dirac operator 
\cite{chirimp} using the implicitly restarted Arnoldi method \cite{arnoldi}.
This is done for both, periodic and anti-periodic temporal b.c. 
for the fermions (the spatial b.c.~are always 
periodic). The chirally improved Dirac operator is not exactly chiral but a 
good approximation of an exactly chiral lattice Dirac operator. The 
topological zero-modes of the continuum manifest themselves as small 
real modes. It can be shown \cite{topsusc}
that although the eigenvalues are not exactly
zero they give rise to the correct topological susceptibility 
through the index theorem.
Although the small real eigenvalues do not vanish exactly we will
refer to the corresponding modes as zero-modes in the following.
From a larger set of configurations we select those configurations
which have a single zero-mode for both b.c., 
i.e.~are in the sector with topological charge $Q = \pm 1$. The restriction to
the $|Q| = 1$ sector is necessary to avoid mixing between 
multiple zero-modes that occurs in higher sectors. For our configurations 
with $|Q|=1$ we find that for both, periodic and anti-periodic 
boundary conditions the zero-mode is always dominated by a single lump.
The position of the lump, however, can be different for the two choices
of the boundary conditions. 

The ensemble in the high temperature phase 
has a non-vanishing Polyakov loop. We will use this ensemble to study the
influence of the expectation value of the Polyakov loop on the zero-modes. We 
divide the deconfining ensemble into sub-ensembles with complex respectively
real Polyakov loop containing 64 (25) configurations. We will refer 
to these subensembles as the complex respectively the real sector.

\begin{figure}[t]
\epsfig{file=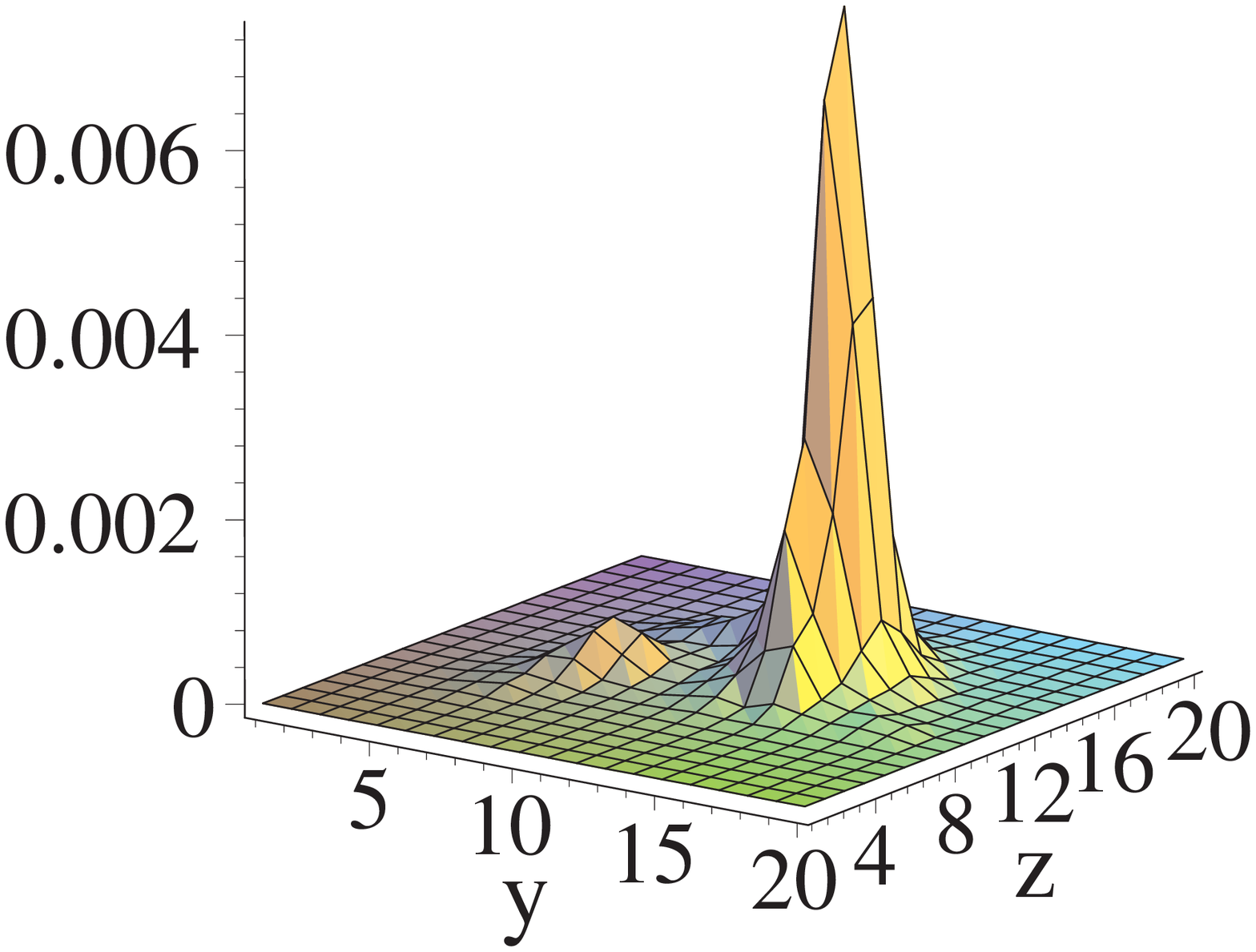,width=4.5cm}
\hspace{-6mm} 
\epsfig{file=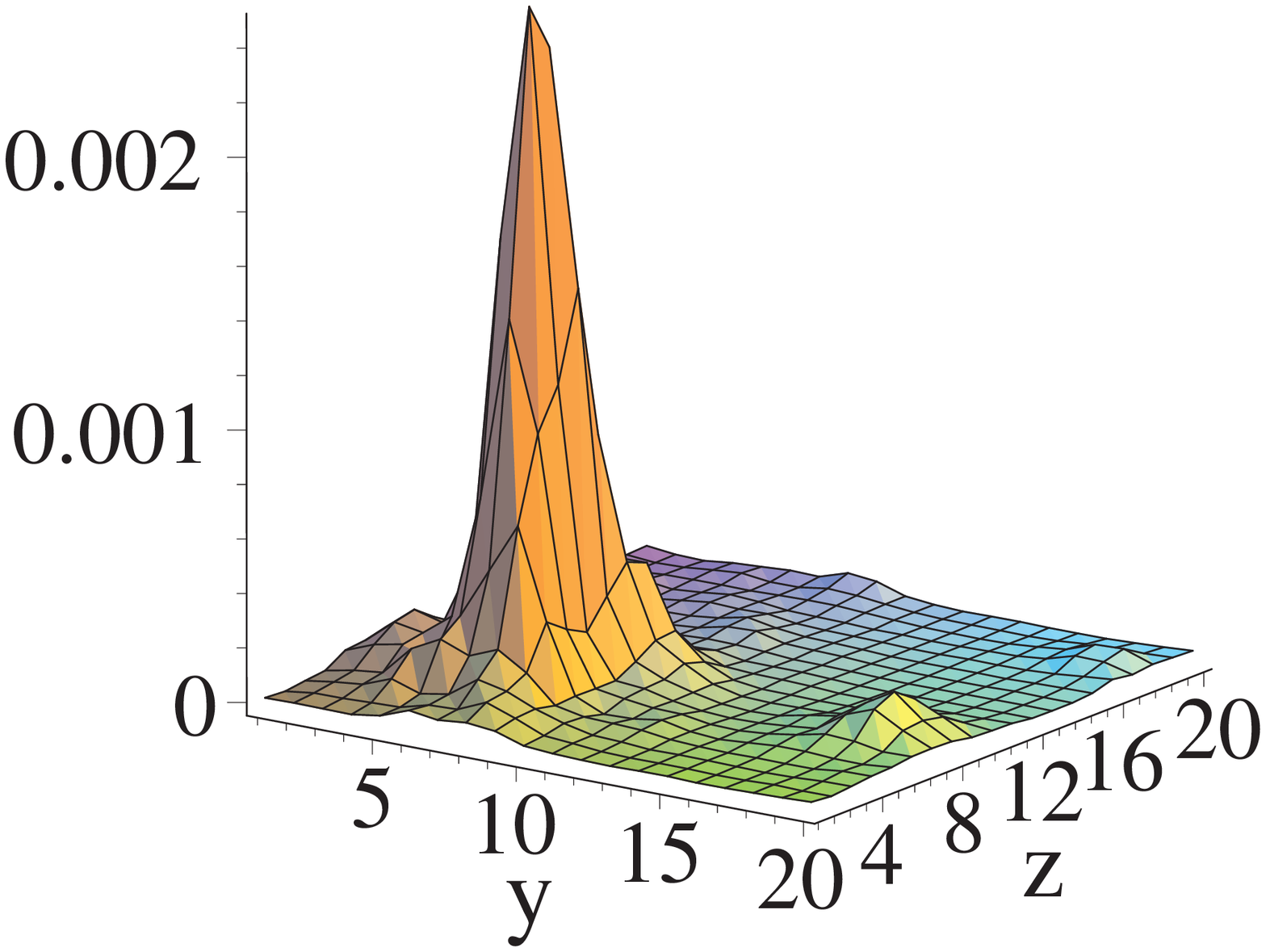,width=4.5cm}
\vskip2mm
\caption{Scalar density of the zero-mode for anti-periodic temporal 
b.c.~(l.h.s.) and periodic temporal b.c.~(r.h.s.). 
We show $y$-$z$ slices for a configuration below $T_c$.
\label{rhoplot}}
\end{figure}

In order to analyze the localization of the zero modes we form the
scalar density $\rho(x) = \sum_{c,d} |\psi(x)_{c,d}|^2$ 
from the zero-modes $\psi(x)_{c,d}$ by summing over the color and Dirac 
indices $c,d$. The resulting density $\rho$ is gauge invariant. 
In Fig.~\ref{rhoplot} we show $y$-$z$ slices of the scalar density 
and compare the anti-periodic case (l.h.s.) to the periodic case (r.h.s.).
The plots are for a configuration in the confining phase. 

Fig.~\ref{rhoplot} clearly demonstrates that the zero-modes with periodic
and anti-periodic b.c.~are located at different space-time
points. For anti-periodic b.c.~$\rho$ has its maximum at 
$(t,x,y,z) = (2,9,13,13)$ while for periodic b.c.~it is located at 
(1,9,6,7). The figure shows the $y$-$z$ slices through the maxima, 
i.e.~the $t=2,x=9$ slice for anti-periodic b.c., respectively 
the $t=1,x=9$ slice for the periodic case. The maxima are separated 
by 9.3 lattice spacings which corresponds to 1.1 fm. It is also obvious
from the figure that the two lumps have different height and width. 
This is not what is expected for the zero-mode of a classical
caloron but agrees well with the behavior predicted for the 
KvB solutions. When scanning through our ensemble 
we have detected many configurations showing such behavior.

Having found instances of configurations resembling KvB 
solutions we now turn to a more quantitative analysis. One interesting
observable was already addressed, the distance between the maximum of 
$\rho_{A}$ and the maximum of $\rho_{P}$. The subscript $A$ refers to 
$\rho$ computed from the zero mode with anti-periodic b.c, while the
subscript $P$ is used for $\rho$ computed from the periodic zero-mode.

A related observable is the overlap of the support of the two lumps. 
The support is defined as the function $\theta(x)$
which has $\theta(x) = 1$ for the lattice points $x$ carrying the 
lump in $\rho$ and $\theta(x) = 0$ for all other points. 
An open question is how to identify the
lump. For a narrow lump the support is small while a broad lump
has a much larger support. 

We will base our definition of the support on
the so-called inverse participation ratio 
$I = V \sum_x \rho(x)^2$,
where $V$ is the total number of points in our lattice. Note that we use 
normalized eigenvectors such that $\sum_x \rho(x) = 1$. It is easy to see that
for an entirely localized state ($\rho(x) = 1$ for only one 
lattice point and 0 else) the
inverse participation ratio is $I = V$, while for a completely
delocalized state ($\rho(x) = 1/V$ for all lattice points)
$I = 1$. The inverse participation ratio is
a convenient measure for the localization, assuming large values for 
localized states and small values for spread out states. 
 
Using $I$ we now define a number $N = V/(32 I)$ and use this to 
determine the support by setting $\theta(x) = 1$ for those $N$ lattice 
points $x$ which carry the $N$ largest values of $\rho$ and $\theta(x) = 0$ 
for all other points. The normalization of $N$ was chosen such that if the 
lump was a 4-D Gaussian the support would consist of the 4-volume 
carrying the points inside the half-width of the Gaussian. Note that
$N$ is the size of the support, i.e.~$N = \sum_x \theta(x)$.

With our definition of the support $\theta$ for
both the anti-periodic ($\theta_A$) and periodic ($\theta_P$)
lumps we can define the 
overlap function $C$ of the two lumps as $C(x) = \theta_P(x)\theta_A(x)$.
From this we compute the relative overlap 
$R = 2/(N_A + N_P) \sum_x C(x)$.
The relative overlap $R$ is a real number ranging from 0.0 to 1.0. If the 
lump in $\rho_A$ and the lump in $\rho_P$ sit on top of each other 
(at least the core up to about the half-width) and have the same size 
then $R = 1.0$. If the two lumps are entirely separated one has 
$R = 0.0$. Values in between give the ratio of the common volume of the 
two supports to the average size of the 
two supports. Note that even if the two lumps sit on top of each other,
but have different sizes $N_A, N_P$ one has $R < 1.0$.
\begin{figure}[t]
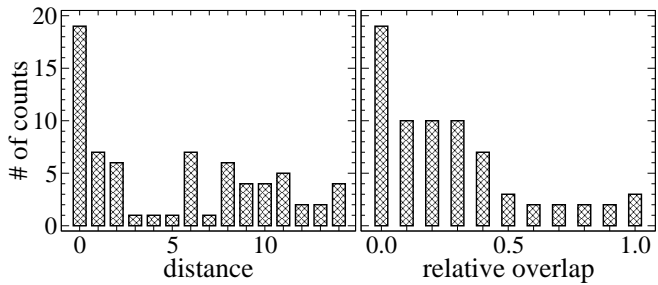

\epsfig{file=disthisto6x20b820.eps,height=3.7cm,clip}
\hspace{-2mm}
\epsfig{file=overlaphisto6x20b820.eps,height=3.7cm,clip}
\vskip2mm
\caption{Histograms for the distance of the maxima (l.h.s.)
and the relative overlap of the lumps for different boundary conditions
(r.h.s.). The data are for the ensemble below $T_c$.
\label{histo6x20b820}}
\end{figure}

In Fig.~\ref{histo6x20b820} we show histograms for the distance 
between the maxima in $\rho_A$ and $\rho_P$ (l.h.s.)~and histograms 
for the relative overlap (r.h.s). 
The distances were rounded to the nearest integer 
while the relative overlap $R$ was rounded to the nearest multiple of $0.1$.
The data shown are for the ensemble below $T_c$. 
When looking at the l.h.s.~plot one finds that for many configurations
the maxima of the two lumps sit on top of each other. 
However, the distribution shows also that for 50\% of the 
configurations the two lumps are separated by at least 6 lattice spacings
which corresponds to 0.7 fm, almost twice the typical instanton size. 
This indicates that for a large fraction of the ensemble the zero-modes
with different b.c.~see different lumps.

Also from the histograms for the relative overlap (r.h.s.~plot in 
Fig.~\ref{histo6x20b820}) one comes to the same conclusion. Only for a 
small fraction of the configurations one finds that the lumps seen by 
anti-periodic and periodic zero-modes coincide and for many cases they do
not overlap at all. A comparison of the two plots in Fig.~\ref{histo6x20b820}
shows that although many pairs of lumps may have the same position of 
their maximum they still differ in size such that their relative overlap 
is smaller than 1. 

In order to further study the evidence for KvB solutions 
we now turn to the ensemble in the high-temperature phase. For these
configurations the theory is deconfined and the Polyakov loop has a
non-vanishing expectation value. As discussed above the 
phase of the Polyakov loop plays an important role for the properties
of the KvB solutions and one expects significant differences
between the complex and the real sector.

\begin{figure}[t]
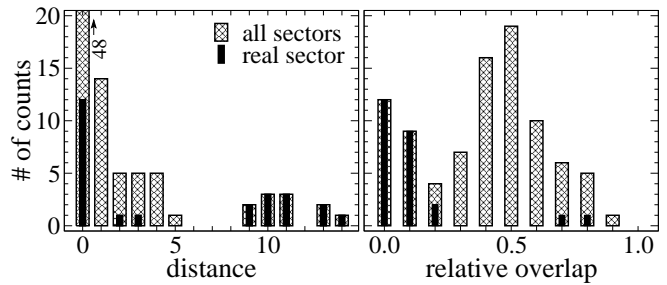

\epsfig{file=disthisto6x20b845.eps,height=3.7cm,clip}
\hspace{-2mm}
\epsfig{file=overlaphisto6x20b845.eps,height=3.7cm,clip}
\vskip2mm
\caption{Same as Fig.~\protect{\ref{histo6x20b820}} but for 
the ensemble above $T_c$.
\label{histo6x20b845}}
\end{figure}

Let us begin our discussion of the deconfined phase by again looking
at the histograms for the distance between the peaks of $\rho$ and
their overlap. Fig.~\ref{histo6x20b845} is equivalent to 
Fig.~\ref{histo6x20b820} but now shows data for the ensemble 
above $T_c$. In addition to showing the data for the full ensemble
we also display the results for the subensemble in the real sector. These 
data are represented by thinner, black bars.

When inspecting the histograms for the distance between the peaks (l.h.s.~plot)
we find that again a certain amount of configurations shows rather large
distances but the fraction of these configurations is smaller. 12\%
of the ensemble show distances of at least 7 lattice spacings 
(again corresponding to $\sim 0.7$ fm). Note that these configurations are 
entirely dominated by the real sector. This is exactly 
as pointed out in our summary 
of the KvB zero-modes where we discussed that in the complex sector the 
two zero-modes sit on the same lump.

Another effect is evident from the two plots in Fig.~\ref{histo6x20b845}.
Although more than half of the configurations have the center of the two 
lumps at exactly the same lattice point we do not see many configurations
where the lumps have a large overlap. The distribution is instead peaked at 
an overlap of 0.5. The reason for this behavior is the difference in size
between the two lumps. One of them is much narrower than the other one 
and thus their relative overlap is considerably smaller than 1. 

\begin{figure}[t]
\centerline{\epsfig{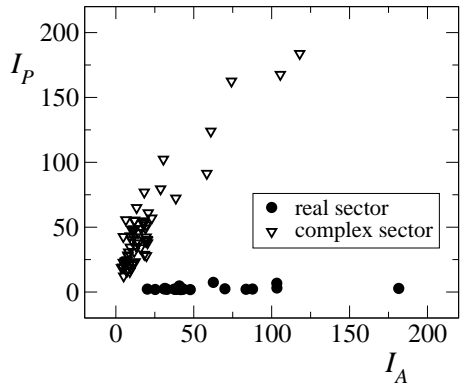}}
\vskip2mm
\caption{Comparison of the inverse participation ratio of the zero mode
for periodic ($I_P$) and anti-periodic 
($I_A$) b.c. The data are from the 
ensemble above $T_c$. 
\label{iprratio}}
\end{figure}

We directly study the phenomenon of the large difference of the sizes of 
the two lumps by comparing the inverse participation ratio $I$
of the zero-mode for anti-periodic and periodic b.c.~($I_A, I_P$).
Fig.~\ref{iprratio} shows a scatter plot of our results with the data
in the real sector represented by filled circles and open triangles are
used for the complex sector. In the real sector one finds that the 
anti-periodic inverse participation ratio (ranging from $I_A \sim 25$ up to 
$I_A \sim 180$) is much larger than its periodic counterpart ($I_P \sim$ 2-5). 
For the complex sector the situation is similar but
the role of the boundary conditions is reversed. However, 
the effect is not as dramatic as in the real sector. 
For not too localized modes ($I_P$ up to 70) one finds that 
the inverse participation ratio for the periodic b.c.~is about 3-times
larger than for the anti-periodic case. The qualitative pattern for the 
different sectors of the Polyakov loop matches 
the predictions for KvB solutions discussed in our summary of the 
properties of the KvB zero-modes.

We remark that we have performed an equivalent analysis also on a 
$16^4$ lattice at $a = 0.094$ fm. Although this is a typical lattice 
for zero-temperature calculations we still find a substantial amount of 
configurations with zero-modes showing the characteristic properties
of KvB solutions. This observation might account for the failure 
of identifying the profile of classical instanton zero-modes in
quenched lattice configurations \cite{horvath2}.

In this letter
we have studied zero-modes of the Dirac operator for gauge 
configurations with topological charge $\pm 1$ 
for temperatures slightly below $T_c$ and a second ensemble in 
the deconfined phase above $T_c$. 
Analyzing the localization of the zero-modes we  
observed large differences when comparing temporal anti-periodic and periodic
b.c.~for the fermions. For a substantial portion of the 
ensemble we found that the zero-mode is located at different positions for
different b.c.~and also the width of the lumps can differ
substantially. For the ensemble in the deconfined phase
we observed that the behavior also depends on the phase 
of the Polyakov loop. The 
emerging picture for the zero-modes is qualitatively well described 
by the properties of the zero-modes of Kraan - van Baal solutions. 
Our results indicate that the picture of topological excitations
being single lumps might be too simple and a substructure resembling
the constituent monopoles of Kraan - van Baal solutions is evident.
\\
\\
{\bf Acknowldegements:} 
I would like to thank Meinulf G\"ockeler, Michael M\"uller-Preussker,
Paul Rakow, Stefan Schaefer, Andreas Sch\"afer, Christian Weiss 
and in particular Pierre van Baal
for discussions and the Austrian Academy of Science for support 
(APART 654). The calculations were done on the Hitachi SR8000
at the Leibniz Rechenzentrum in Munich and I thank the LRZ staff for
training and support.

\end{document}